\documentclass[twocolumn,secnumarabic,preprintnumbers,amsmath,amssymb,10pt,a4paper,superscriptaddress,aps,pre,showpacs]{revtex4-1}
\usepackage{geometry}
\usepackage{graphicx}
\usepackage{dcolumn}
\usepackage{bm}
\usepackage{mathtools}

\renewcommand{\vec}[1]{\mathbf{#1}}
\newcommand{\e}[1]{\mathcal{E}^{(#1)}}
\newcommand{\order}[1]{\mathcal{O}\left(#1\right)}
\newcommand{\equals}{\!=\!}
\newcommand{\great}{\!>\!}
\renewcommand{\Re}{{\rm Re}}
\newcommand{\cL}{\mathcal{L}}
\newcommand{\dd}{{\rm d}}
\newcommand{\dt}{{\rm d}t}
\newcommand{\dtau}{{\rm d}\tau}
\newcommand{\id}{{\rm 1\!\!I}}
\newcommand{\avg}[1]{\left\langle{#1}\right\rangle}

\usepackage{color}

\begin{document}

\title{When the mean is not enough: Calculating fixation time distributions in birth-death processes}

\author{Peter Ashcroft}
\email{peter.ashcroft@manchester.ac.uk }
\affiliation{Theoretical Physics, School of Physics and Astronomy, The University of Manchester, Manchester M13 9PL, United Kingdom}

\author{Arne Traulsen}
\email{traulsen@evolbio.mpg.de}
\affiliation{Max-Planck-Institute for Evolutionary Biology, August-Thienemann-Str. 2, 24306 Pl\"on, Germany}

\author{Tobias Galla}
\email{tobias.galla@manchester.ac.uk}
\affiliation{Theoretical Physics, School of Physics and Astronomy, The University of Manchester, Manchester M13 9PL, United Kingdom}

\begin{abstract}
Studies of fixation dynamics in Markov processes predominantly focus on the mean time to absorption. This may be inadequate if the distribution is broad and skewed. We compute the distribution of fixation times in one-step birth-death processes with two absorbing states. These are expressed in terms of the spectrum of the process, and we provide different representations as forward-only processes in eigenspace. These allow efficient sampling of fixation time distributions. As an application we study evolutionary game dynamics, where invading mutants can reach fixation or go extinct. We also highlight the median fixation time as a possible analog of mixing times in systems with small mutation rates and no absorbing states, whereas the mean fixation time has no such interpretation.
\end{abstract}


\maketitle

\section{Introduction} \label{sec:Introduction}
If a group of mutants is introduced into a population of wild-type individuals, how long does it take for the population to reach a homogeneous state? This is a typical question asked in population genetics and evolutionary biology. It can be addressed using the theory of stochastic processes and approaches from statistical physics.

In its simplest form, evolutionary dynamics is modeled as a Markov process in a population of $N$ individuals of two types \cite{nowak2006evolutionary}. Fixation occurs when the process arrives at one of its absorbing states. As the time to fixation is itself a random variable, only the computation of the distribution of fixation times provides a complete answer to our opening question. The majority of existing studies avoid this mathematical challenge and instead focus on calculating only the mean fixation time \cite{kimura1980average,antal2006fixation,altrock2009fixation}. While the first moment can provide a good indication of the outcome in some circumstances, this approach can be insufficient when the distribution of fixation times is broad \cite{dingli:CCY:2007,altrock2011stability}. As we show in this paper, such scenarios arise in examples from evolutionary game theory.

Although the master equation describing the birth-death dynamics is linear, calculating fixation time distributions is more intricate than one may initially think. Nested expressions for all moments of fixation times are known \cite{goel1974stochastic,ewens,altrock2011stability} and from these the distribution can in principle be constructed recursively up to arbitrary precision. However, this approach does not provide a simple closed-form solution or a means of efficiently sampling from the arrival time distribution.

An alternative approach is to diagonalize the linear operator of the master equation and to carry out the analysis in eigenspace. This leads to a theorem attributed to Karlin and McGregor \cite{karlin1959coincidence,keilson1979markov}, which states that arrival times can be written as the sum of independent, exponentially distributed random variables parametrized by the eigenvalues of the master equation. Alternatively stated, the arrival time distribution is given by the convolution of the exponential distributions from which the independent random numbers are sampled, i.e., it is a phase-type distribution \cite{ocinneide1991phase}. This theorem has been discussed in numerous sources in the probability theory literature \cite{keilson1979markov, brown1987identifying, fill1991time, brown1999interlacing, fill2009passage, diaconis2009times,fill2009hitting, miclo2010absorption, gong2012hitting}. The discussion of these matters is usually very terse, and not easily accessible to physicists or researchers in adjacent disciplines. Researchers in the theoretical biosciences are only recently beginning to use these ideas for the purpose of model reduction \cite{barrio2013reduction, leier2014exact}. Existing results are limited to specific initial conditions and types of birth-death chains, and a clear understanding of the analysis in eigenspace is lacking.

We consider one-step birth-death processes with two absorbing states and a general initial condition. This model describes the invasion (or extinction) of a number of mutants in a wild-type population. It is a generalization of existing studies, which focus on a single absorbing boundary and a specific initial condition \cite{karlin1959coincidence}. The purpose of our work is to provide a systematic and explicit closed-form solution for fixation time distributions, and a physical interpretation of different representations in eigenspace \cite{brown1999interlacing,fill2009hitting,miclo2010absorption}. By generalizing the Karlin-McGregor result to account for arbitrary initial conditions and multiple absorbing states, we show that these different representations can be traced back to one common origin. To generate samples from the fixation time distribution it is sufficient to simulate forward-only processes in eigenspace. This provides effective model-reduction schemes.

We use our results to relate fixation processes to the equilibration dynamics of evolutionary systems with mutation (and hence with no absorbing states). These equilibration processes are often characterized by the so-called mixing time, the time it takes the system to come within a set distance of its stationary distribution \cite{levin2009markov,black2012mixing}. We thus discuss an appropriate analog of the fixation time in the limit of small mutation rates, and outline the limitations of this relation.

The paper is organized as follows: In Sec.~\ref{sec:Model} we describe the model used and the principal idea behind the work. In Sec.~\ref{sec:Analysis} we outline the mathematical procedure that takes us from the model to the arrival time distributions. Details of the calculations can be found in the Appendixes. In Sec.~\ref{sec:EvoGames} we apply our results to specific examples of evolutionary games, confirming both the accuracy and efficiency of our method. In Sec.~\ref{sec:Equilibration} we discuss the relationship between the fixation time and mixing time in the limit of small mutation rates, before drawing conclusions in Sec.~\ref{sec:Conclusion}.

\section{Model definition and main idea} \label{sec:Model}
\begin{figure}[t]
\centering
\includegraphics[width=.95\columnwidth]{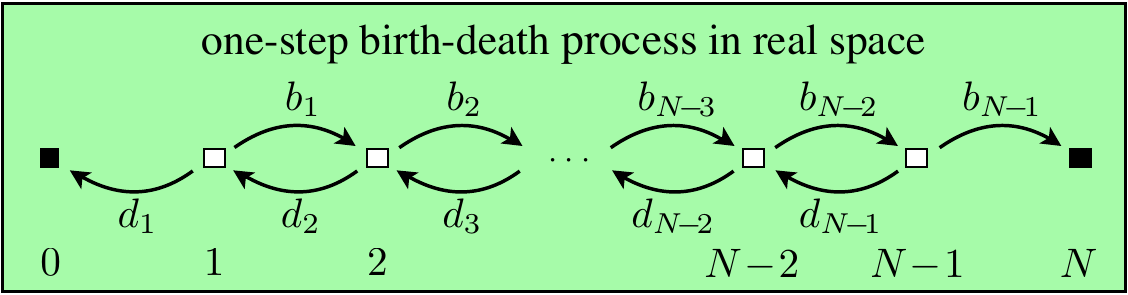}
\caption{(Color online) One-step birth-death process in a population of $N$ individuals. The variable $i$ denotes the number of invading mutants. The states $i=0$ (extinction) and $i=N$ (fixation) are absorbing. Birth rates are labeled $b_i$ and death rates $d_i$.}
\label{fig:fig1}
\end{figure}

We study a one-step birth-death process with states $i=0,1,\dots, N$ and characterized by the birth and death rates $b_i, d_i$ ($i=1,\dots,N-1$), as illustrated in Fig.~\ref{fig:fig1}. This describes a population of constant size $N$ with $i$ individuals of the mutant type and $N-i$ of the resident wild type \cite{nowak2006evolutionary}. The states $i=0$ and $i=N$ are absorbing in the absence of mutation, and so the dynamics will end at one of these two states eventually. Our objective is to calculate the distribution of arrival times at these absorbing states for a general starting point $i_0$.

\begin{figure}[t]
\centering
\includegraphics[width=.95\columnwidth]{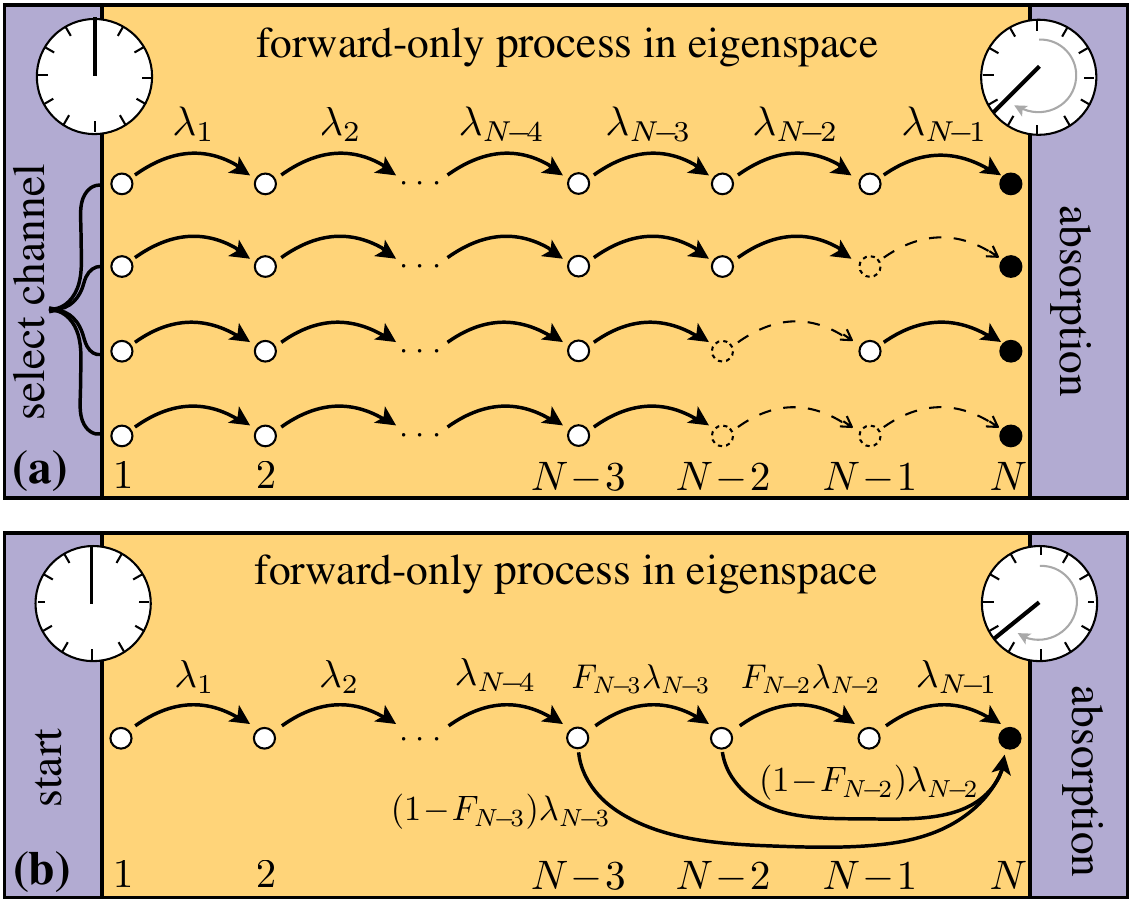}
\caption{(Color online) Different representations of the arrival process in terms of forward-only processes in eigenspace. The $\lambda_\alpha$ are (absolute) eigenvalues of the process in Fig.~\ref{fig:fig1}, each arrow represents an exponential process with the rate indicated.
(a) shows a set of alternative reaction channels. In each run one channel is chosen with appropriate probability. Transitions indicated by dashed arrows are skipped (zero time).
(b) shows a single forward-only chain, in which the final state can be reached directly from some of the intermediate states. Both representations are equivalent and generate samples from the arrival time distribution of the process in Fig.~\ref{fig:fig1}. The case shown here is for arrival at $N$, starting from $i_0=3$ in the original space.}
\label{fig:fig2}
\end{figure}

The outcome of our analysis (detailed below) are the reduced forward-only processes in eigenspace illustrated in Fig.~\ref{fig:fig2}, which have the same arrival time distribution as the original birth-death process. The forward jump rates are determined by the eigenvalues $\lambda_\alpha$ of the original birth-death process.
The schematic in Fig.~\ref{fig:fig2}(a) shows multiple forward-only channels, where dashed arrows indicate that some steps are to be skipped. Figure~\ref{fig:fig2}(b) shows a single forward-only channel. Long arrows indicate direct jumps to the final eigenstate. In both cases, the number of possible paths is determined by the initial condition and the final state ($i=0$ or $i=N$) of the original birth-death process.

Arrival time samples of the original process are generated from Fig.~\ref{fig:fig2}(a) in the following way: One of the channels is chosen with probability determined by the birth-death rates, the initial condition, and the arrival state of the original real-space process. After a channel has been selected, the clock is started and the forward-only process of the channel is traversed. The clock is stopped when the final state in the schematic is reached (absorption).

Arrival time samples are generated from Fig.~\ref{fig:fig2}(b) by traversing the forward-only chain. The quantities $1-F_\alpha$ are the probabilities to reach absorption directly from the intermediate eigenstate $\alpha$. These probabilities are again determined by the birth-death rates, the initial condition and the final state of the original process, and explicit formulas can be found in Appendix~\ref{app:Compute}\,\ref{app:Matching}.
 
\section{Analysis} \label{sec:Analysis}
To derive these results it is convenient to focus on the interior states $i=1,\dots, N-1$ in the birth-death process in Fig.~\ref{fig:fig1}. The probability to be in state $i$ at time $t$, $p_i(t)$, satisfies the equation $\dot {\vec{p}}=\mathbb{A}\cdot\vec{p}$ with formal solution $\vec{p}(t)=\exp(\mathbb{A}t)\cdot\vec{p}(0)$. The matrix $\mathbb{A}$ describes the transient states and has elements $a_{i,i} = -(b_i+d_i)$, $a_{i,i+1}=d_{i+1}$ and $a_{i,i-1}=b_{i-1}$. The initial condition is $p_i(0)=\delta_{i,i_0}$ ($1 \le i_0 \le N-1$). Probability continuously flows into the two absorbing states. Hence all eigenvalues of $\mathbb{A}$ are negative, and so we focus on $-\mathbb{A}$ and its eigenvalues. The matrix exponential can be evaluated by Laplace transformation, and subsequent back transformation allows one to calculate $\dot{P}_{0|i_0}(t)=d_1 p_1(t)$ and $\dot{P}_{N|i_0}(t)=b_{N-1} p_{N-1}(t)$. Mathematical details can be found in Appendix~\ref{app:Calculate}. Up to normalization these are the conditional arrival time distributions at states $0$ and $N$ respectively. For the remainder of the paper we focus only on absorption at state $N$. Analogous expressions for absorption at $0$ can be found in the Appendixes. We find the following expression for the probability flux (per unit time) into state $N$
\begin{equation}
\dot P_{N|i_0}(t) = \frac{B_{i_0}\psi_{i_0}}{\Lambda} E_{N-1} * R_{i_0-1},
\label{eq:gen_result}
\end{equation}
where $B_{i_0}=\prod_{i=i_0}^{N-1}b_i$, $\Lambda=\det(-\mathbb{A})$ and $\psi_{i_0}$ is the determinant of the $(i_0-1)\times(i_0-1)$ top-left sub-matrix of $-\mathbb{A}$ (denoted as $-\mathbb{A}^{(i_0-1)}$). We have introduced $E_{\ell}=\e{\lambda_1} * \cdots * \e{\lambda_\ell}$, where the $\e{\lambda_\alpha}(t)$ are exponential distributions with the eigenvalues of $-\mathbb{A}$, $\lambda_\alpha>0$, as parameters. The symbol $*$ represents a convolution. The object $R_\ell$ is of the form
\begin{equation}
R_\ell = \left(\delta+y_1^{-1}\delta'\right) * \dots * \left(\delta+y_\ell^{-1}\delta'\right),
\label{eq:convolve}
\end{equation}
where $y_\alpha>0$ are the eigenvalues of the sub-matrix $-\mathbb{A}^{(i_0-1)}$. In these expressions $\delta$ is the usual Dirac distribution, and $\delta'$ is its derivative, as described in Appendix~\ref{app:Background}\,\ref{app:Dirac}. We can identify the prefactor $B_{i_0}\psi_{i_0}/\Lambda=\phi_{N|i_0}$ as the probability that the original system gets absorbed in state $N$ (as opposed to $0$).

The effective dynamics shown in Fig.~\ref{fig:fig2} are obtained by evaluating the convolutions in Eq.~\eqref{eq:gen_result}. To illustrate this it is useful to first consider the convolution of the exponential distribution $\e{\lambda}(t)$ with an object of the form $\delta(t) + y^{-1}\delta'(t)$ ($y>0$). In Appendix~\ref{app:Background}\,\ref{app:Conv} we show that
\begin{equation}
\e{\lambda} * \left(\delta+y^{-1}\delta'\right) = \left[\frac{\lambda}{y}\delta(t) + \left(1-\frac{\lambda}{y}\right)\e{\lambda}(t)\right]\!.
\label{eq:expconv}
\end{equation}
Assuming $\lambda/y<1$ (which will be the case throughout our analysis) this describes a convex combination of a point mass at zero and an exponential distribution. For samples, set $t=0$ with probability $\lambda/y$, otherwise draw $t$ from $\e{\lambda}(t)$.

To arrive at the dynamics depicted in Fig.~\ref{fig:fig2}(a), each of the $i_0-1$ terms of the form $\delta+y_\alpha^{-1}\delta'$ in Eq.~\eqref{eq:gen_result} is paired up with a separate exponential, as described in Appendix~\ref{app:Compute}\,\ref{app:Pairing}. This creates a total of $2^{i_0-1}$ possible forward-only channels with up to $i_0-1$ exponential steps skipped, as shown in Fig.~\ref{fig:fig2}(a). To arrive at the dynamics shown in Fig.~\ref{fig:fig2}(b), the $i_0-1$ objects of the form $\delta+y_\alpha^{-1}\delta'$ are successively convoluted with the full chain $E_{N-1}$ from the right, as described in Appendix~\ref{app:Compute}\,\ref{app:Matching}. This leads to $i_0$ channels, in which $0, 1,\dots,i_0-1$ exponential steps are skipped. This can be illustrated as a single forward-only channel, with appropriate rates of jumping from intermediate eigenstates to absorption. By considering these forward-only processes we arrive at closed-form expressions for the arrival time distributions. For arrival at state $N$, the distribution is given by (see Appendix~\ref{app:Compute}\,\ref{app:BottomLine})
\begin{equation}
\dot{P}_{N|i_0}(t) = B_{i_0} \sum_{\alpha=1}^{N-1}\left[ \frac{\prod\limits_{\gamma=1}^{i_0-1} \left(y_\gamma-\lambda_\alpha\right)}{\prod\limits_{\substack{\beta=1\\\beta\ne \alpha}}^{N-1}(\lambda_\beta-\lambda_\alpha)}e^{-\lambda_\alpha t}\right].
\label{eq:bottomline}
\end{equation}

The representation shown in Fig.~\ref{fig:fig2}(a) corresponds to the picture obtained for a restricted set of processes by probabilistic methods in Ref.~\cite{brown1999interlacing}. On the other hand, Fig.~\ref{fig:fig2}(b) reflects the findings of Refs.~\cite{fill2009hitting,miclo2010absorption}, derived from the construction of intertwining processes. Our analysis shows that these different decompositions originate from one common structure, Eq.~\eqref{eq:gen_result}. The explicit schemes in Fig.~\ref{fig:fig2} provide a computational method to generate samples from the arrival time distribution efficiently, for example by carrying out simulations of these forward processes using the Gillespie algorithm \cite{gillespie1977exact}. It is important to keep in mind that the eigenstates shown in Figs.~\ref{fig:fig2}(a) and \ref{fig:fig2}(b) cannot be mapped one to one to the states in Fig.~\ref{fig:fig1}. The equivalence of the real and eigenspace representations only holds on the level of arrival time statistics.

\section{Evolutionary games} \label{sec:EvoGames}
As an application of this theory we now consider examples of evolutionary dynamics with frequency-dependent selection \cite{nowak2006evolutionary,szabo2007evolutionary,frey2010evolutionary}. Such models are used to describe the interaction of invading mutants ($A$) and resident wild-type individuals ($B$). These scenarios can be formulated as evolutionary games \cite{nowak2006evolutionary}. We focus on a $2\times 2$ normal form game,
\begin{equation}
\begin{array}{c|cc}
   & A & B \\ \hline
A & R & S \\
B & T & P
\end{array}~~
\qquad
\begin{array}{ll}
\pi_A(i)=\frac{i-1}{N-1}R+\frac{N-i}{N-1}S, \vspace{0.2cm}\\
\pi_B(i)=\frac{i}{N-1}T+\frac{N-i-1}{N-1}P,
\end{array}
\label{eq:game}
\end{equation}
where $\pi_A(i)$ and $\pi_B(i)$ are the expected payoffs in a population of $i$ individuals of type $A$ and $N-i$ individuals of type $B$. For this example we assume a pairwise comparison process, leading to birth and death rates given by $b_i\equals g[+\Delta\pi(i)] {i(N-i)}/{N}$ and $d_i\equals g[-\Delta\pi(i)] {i(N-i)}/{N}$, respectively, where $\Delta \pi(i)\equals \pi_A(i)-\pi_B(i)$ and $g(z)\equals \left(1 + \beta z\right)/2$. The parameter $\beta\great 0$ is the so-called intensity of selection \cite{,nowak2004emergence,nowak2006evolutionary,pinheiro2012selection}. The parameters $R, S, T, P$ specify the interaction, and we consider three possible scenarios: coexistence ($T\great R$ and $S\great P$, stable heterogeneous population); coordination ($R\great T$ and $P\great S$, unstable heterogeneous population); and the prisoner's dilemma ($T\great R\great P\great S$, stable homogeneous population) \cite{nowak2006evolutionary}.

\begin{figure}
\centering
\includegraphics[width=0.95\columnwidth]{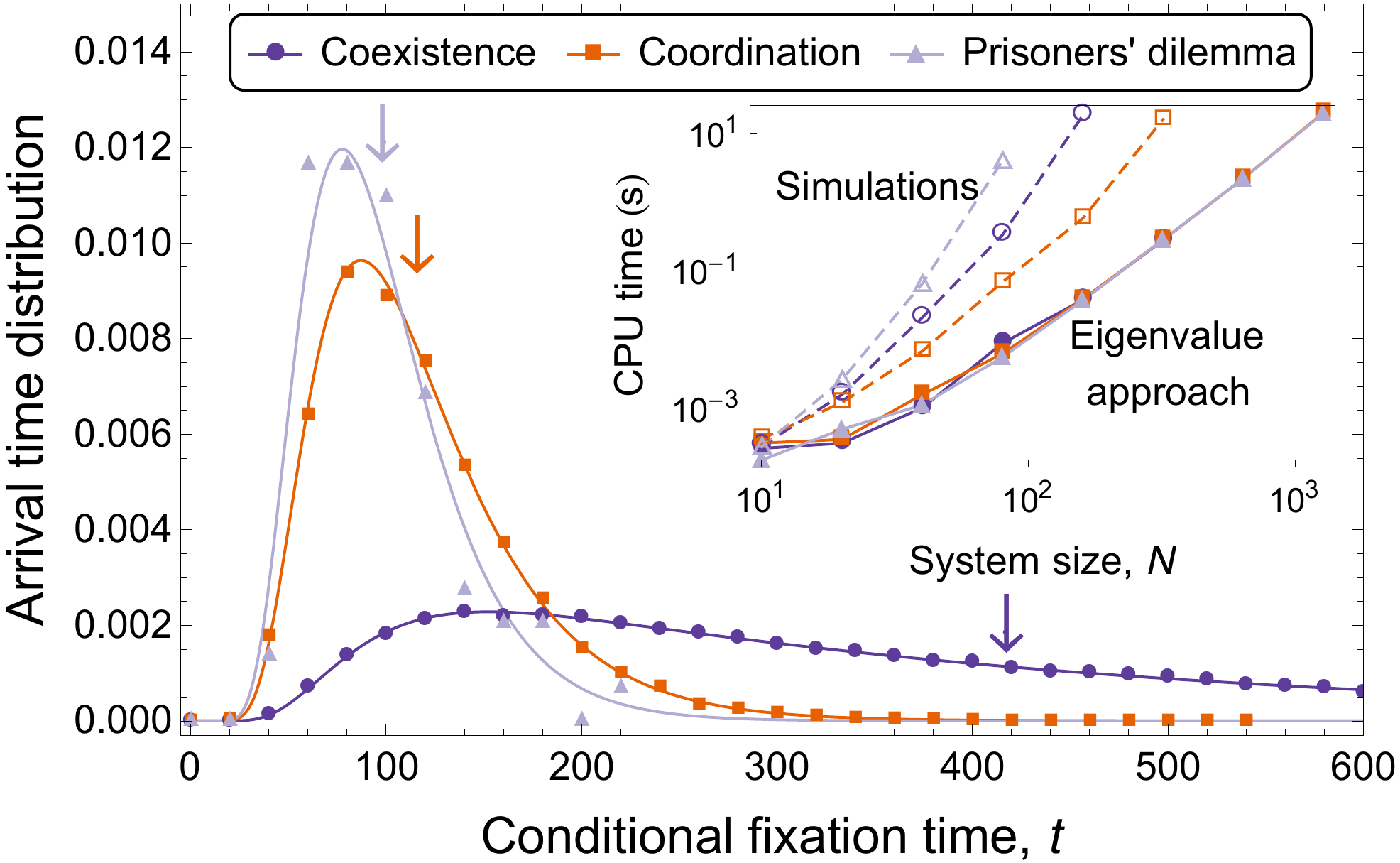}
\caption{(Color online) The conditional fixation time distributions at $i\equals N$ for different evolutionary games. Main panel: Lines show results from the theory [Eq.~\eqref{eq:bottomline}], symbols are from simulations ($10^6$ runs per game). The mean fixation time (arrows) is not a good description of the distribution. Inset: Open markers/dashed lines show computer time needed to obtain arrival time distributions from simulations, full markers/solid lines are for the semi-analytical approach and indicate the polynomial scaling. Direct simulations can only approximate the exact arrival time distribution to a given accuracy (see Appendix~\ref{app:Compute}\,\ref{app:Accuracy} for details). ($N\equals 100$, $i_0\equals 10$, $\beta\equals 0.1$. Coexistence game: $R\equals P\equals 1.0,~S\equals T\equals 1.5$; Coordination game: $R\equals P\equals 1.5,~S\equals T\equals 1.0$; prisoner's dilemma: $R\equals -S\equals 0.5$, $T\equals 1.0$, $P\equals 0.0$).}
\label{fig:fig3}
\end{figure}

As shown in Fig.~\ref{fig:fig3}, arrival time distributions can be broad and skewed, such that the mean fixation time contains only limited information. Figure~\ref{fig:fig3} also demonstrates the computational benefits of our formalism. Evaluating the distribution in Eq.~\eqref{eq:bottomline} requires $\order{N^3}$ operations as the spectra of $-\mathbb{A}$ and $-\mathbb{A}^{(i_0-1)}$ are needed. Generating arrival time distributions from simulation of the original birth-death process instead can take exponentially long. For the coordination game and the prisoner's dilemma, fixation at $N$ is rare and the bottleneck of direct simulations is the limited sampling. For coexistence games, fixation times are exponentially long in $N$. This impedes accurate simulations. Direct numerical integration of the master equation would suffer from the same problem.

\section{Equilibration processes in systems with mutation} \label{sec:Equilibration}
We now consider birth-death processes without absorbing states by adding mutation occurring at a rate $u\ll1$, such that $b_0=\order{u}$ and $d_N=\order{u}$. All other transition rates depicted in Fig.~\ref{fig:fig1} are $\order{u^0}$ and are only affected at sub-leading order by $u$ (exact transition rates can be found in Appendix~\ref{app:Equilibration}\,\ref{app:Mutation}). The timescale of the dynamics is characterized by the so-called `mixing time', $t_{\rm mix}$. This is the time taken for the probability distribution, $\vec{P}(t)$, to come within a specified distance of the stationary distribution $\vec{P}^{\rm st}$, i.e., $t_{\rm mix}$ is the first time at which $d[\vec{P}(t_{\rm mix}),\vec{P}^{\rm st}]=\varepsilon$. The distance between distributions $\vec{P}$ and $\vec{Q}$ commonly used in this context is $d(\vec{P},\vec{Q})=\sum_{i=0}^{N} |P_i- Q_i|/2$ with $\varepsilon=1/2$ \cite{levin2009markov,black2012mixing}. Using our results we can determine if and when there is a correspondence between the mixing time and the fixation time.

For very small mutation rates ($0<uN\ll 1$) the stationary distribution is of the form $P_i^{\rm st}\approx (1-\sigma) \delta_{i,0}+\sigma\delta_{i,N}$, where $\sigma=\order{u^0}$ (see Appendix~\ref{app:Equilibration}\,\ref{app:Mutation}). 
In the strict absence of mutation ($u=0$) the system reaches fixation, so its terminal distribution, $\vec{\Phi}$, is of the form $\Phi_{i|i_0}=(1-\phi_{N|i_0}) \delta_{i,0}+\phi_{N|i_0}\delta_{i,N}$. It is clear that these two distributions are different; $\vec{P}^{\rm st}$ in systems with $u>0$ is independent of the initial condition. Thus there is no obvious connection between fixation times and mixing times in the limit $u\to 0$.

\begin{figure}[t]
\centering
\includegraphics[width=0.95\columnwidth]{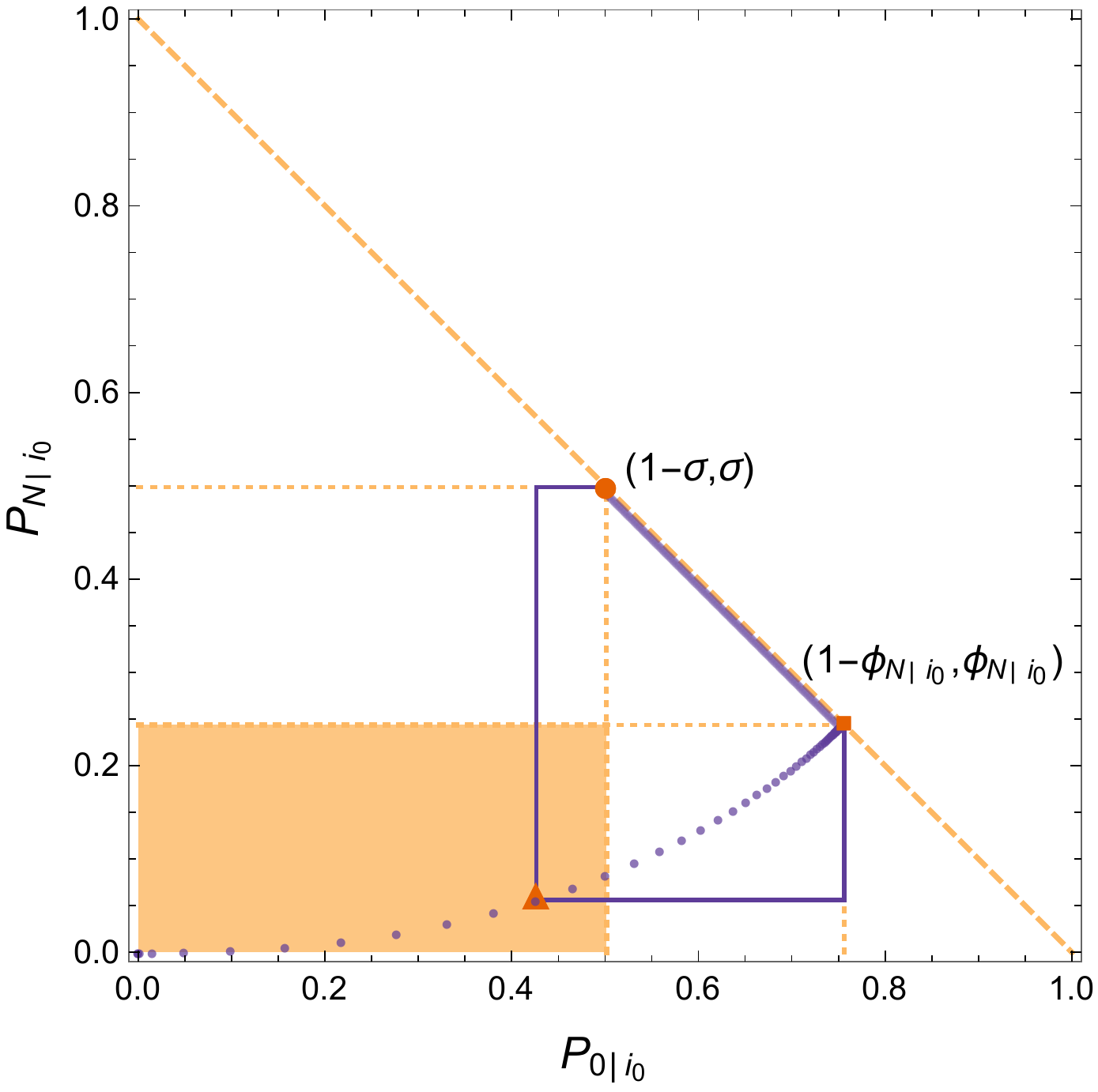}
\caption{(Color online) Approach to the stationary distribution for a system with small mutation rates. Reaction rates are given in Eq.~\eqref{eq:app:mutation}. Dots show the trajectory of $[P^{(u)}_0(t),P^{(u)}_N(t)]$ (found from numerical integration of the master equation with mutation). The probability quickly approaches the fixation distribution, $\vec{\Phi}$, before slowly leaking to the stationary state, $\vec{P}^{\rm st}$. For $0<i_0<N$, the trajectory starts at $(0,0)$ and leaves the shaded area at time $t^*$. For any point inside the shaded area, the distance to the points $(1-\phi_{N|i_0},\phi_{N|i_0})$ and $(1-\sigma,\sigma)$ in our metric (solid lines) are equal. Payoff parameters are $R=P=1.25$ and $S=T=0.75$, which corresponds to a coordination game. Remaining parameters are $\beta=0.1$, $N=100$, $i_0=35$, and $u=10^{-5}$.}
\label{fig:fig4}
\end{figure}

However, equilibration in many systems with rare mutations is a two-step process; the system first reaches an intermediate distribution that is dependent on the initial condition, before `leaking' on a longer timescale into the final stationary state \cite{black2012mixing}. This is analogous to the quasi-stationary distribution before reaching absorption in systems without mutation \cite{diaconis2009times}. Our analysis suggests that this intermediate distribution of systems with $0<uN\ll 1$ is close to the terminal distribution $\vec{\Phi}$ of the system with $u=0$, and that both systems initially evolve along similar trajectories (see Appendix~\ref{app:Equilibration}\,\ref{app:Mutation}). This can be seen in Fig.~\ref{fig:fig4} where we represent the probability distribution as single points in the $(P_0,P_N)$-plane. The distribution first approaches the terminal distribution $\vec{\Phi}$ before slowly converging to the stationary distribution. The most appropriate analog of fixation times in systems with small mutation rates is thus the time to reach this intermediate distribution, not the time to stationarity (mixing time).

\begin{figure}[t]
\centering
\includegraphics[width=0.95\columnwidth]{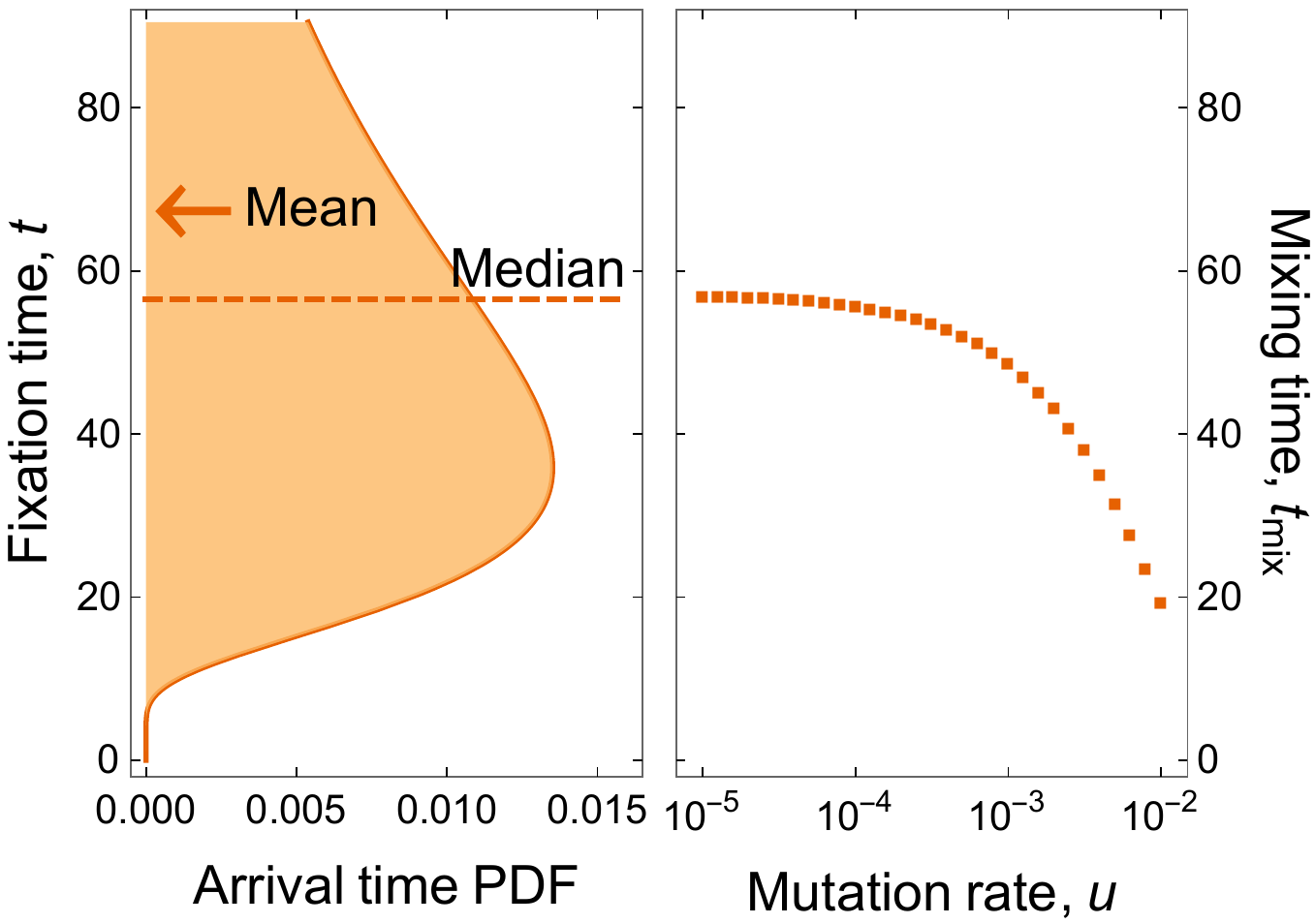}
\caption{(Color online) Correspondence of mixing time and median fixation time for small mutation rates. (a) shows the unconditional fixation time distribution for the coordination game ($u=0$). (b) depicts the mixing time for $u>0$ and $\varepsilon=1/2$ (from numerical integration of the master equation \cite{black2012mixing}). Reaction rates and parameters are as in Fig.~\ref{fig:fig4}.}
\label{fig:fig5}
\end{figure}

\begin{figure}[t]
\centering
\includegraphics[width=0.95\columnwidth]{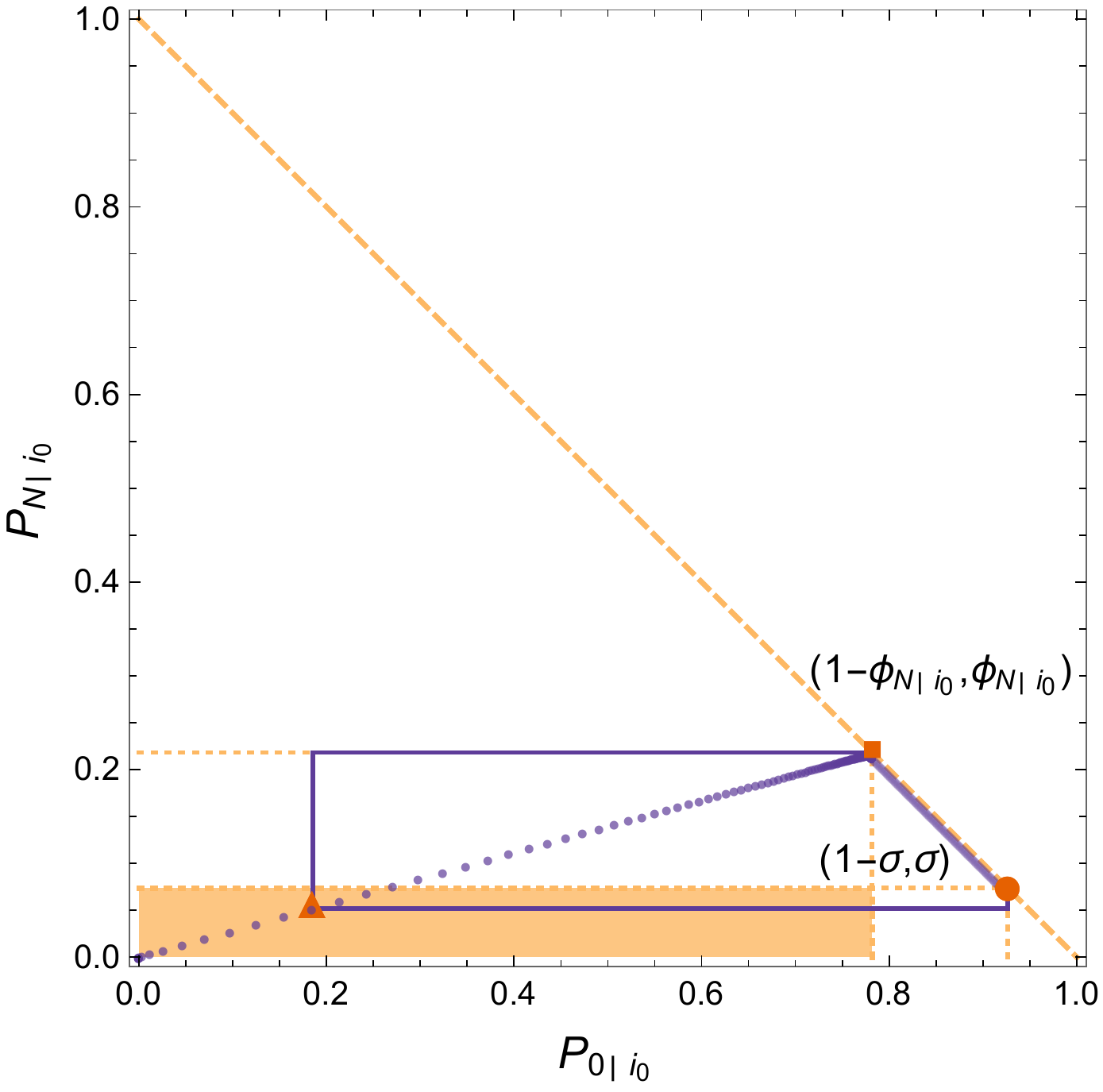}
\caption{(Color online) This figure is as described in Fig.~\ref{fig:fig4}, except the dynamics now follow the prisoner's dilemma scenario. At time $t^*$ we have $d(t^*)\approx 0.65$, which means the mixing time and the median fixation time do not correspond in this example. Parameters are $R=-S=0.5$, $T=1.0$, $P=0.0$, $\beta=0.025$, $N=100$, $i_0=50$, and $u=10^{-5}$.}
\label{fig:fig6}
\end{figure}

The mixing time can related to the fixation time in a different way:
The probability distribution initially satisfies $d[\vec{P}(t),\vec{P}^{\rm st}] = d[\vec{P}(t),\vec{\Phi}]$, which is a consequence of the distance measure used (solid lines in Fig.~\ref{fig:fig4}). This equivalence holds while $P_0(t)<1-\sigma$ and $P_N(t)<\sigma$ (see Appendix~\ref{app:Equilibration}\,\ref{app:Mutation}). Prior to the first time that this condition is violated, $\vec{P}(t)$ is approximately the same as in the system without mutation where we have ${\rm Pr}(t_{\rm fix}<t) = 1-d[\vec{P}(t),\vec{\Phi}]$ for the cumulative fixation time distribution. The condition $d[\vec{P}(t),\vec{\Phi}]=1/2$ therefore translates into the time at which half of the samples have reached fixation, and provided the above conditions hold there is a correspondence between the mixing time and the {\em median} fixation time. This is illustrated in Fig.~\ref{fig:fig5} where we consider the example of a coordination game with a symmetric stationary distribution \cite{black2012mixing}. If $\vec{P}^{\rm st}$ is asymmetric and $\sigma \ll (1-\sigma)$ (or vice versa), then the condition $P_N(t)<\sigma$ is violated after a short period of time. In such cases the above correspondence may not hold for the median fixation time, but only for a lower percentile. This is the case for the example shown in Fig.~\ref{fig:fig6}.

\section{Conclusion} \label{sec:Conclusion}
In summary, we have constructed eigenspace representations that capture the full arrival time statistics of one-step birth-death processes. The mapping into eigenspace has a clear interpretation as forward-only exponential processes. Sampling of the original arrival time distributions reduces to simulating these forward-only processes, or equivalently evaluating a finite sum of exponential random variables, turning our results into an effective tool for model reduction. The compact structure of the forward-only processes allows us to derive exact, closed-form expressions for the arrival time distributions of the original process in terms of its spectrum. As we have demonstrated, the numerical evaluation of these expressions is an efficient polynomial-time method to obtain full arrival time statistics. We have also established a link between equilibration times in systems with small mutation rates and the median fixation time in absence of mutation in some circumstances. Our work advances the understanding of the mathematical structure underpinning the dynamics of fixation. We have placed existing representations for simpler cases into a wider and more coherent context \cite{brown1999interlacing,fill2009hitting,miclo2010absorption}. Nevertheless, there are fundamental open questions. Claims of probabilistic interpretations of Karlin and McGregor's theorem have been made \cite{fill2009passage,diaconis2009times}, but in our view this picture is still incomplete. We would argue that a full probabilistic interpretation of the representations in eigenspace is only reached when each time step of the forward-only process can be constructed directly and uniquely from realizations of the original process alone. Whether or not this is possible is unclear.

\appendix
\section{Mathematical background material} \label{app:Background}
\subsection{Dirac distribution and its derivative} \label{app:Dirac}
The Dirac-$\delta$ distribution has support $\{0\}$. It can be written in its Fourier representation, $\delta(t)=\int_{-\infty}^\infty d\omega\, e^{i\omega t}$. The distributional derivative, $\delta'$, can be conveniently defined by its Fourier transform as well \cite{risken}. It has the form
\begin{equation}
\delta'(t) = \int_{-\infty}^\infty  (i\omega)e^{i\omega t}\,\dd\omega.
\end{equation}
When this is convoluted with a test function, $f(t)$, with infinite support one obtains (after integration by parts)
\begin{equation}
\int_{-\infty}^{\infty} \delta'(t-\tau) f(\tau)\,\dtau = f'(t).
\end{equation}
If a test function, $g(t)$, has finite support, say $t\ge0$, then one finds
\begin{equation}
\int_0^{\infty} \delta'(t-\tau) g(\tau)\,\dtau = g'(t)+g(0)\delta(t).
\end{equation}

\subsection{Laplace transform of an exponential distribution}
We frequently use the Laplace transform of an exponential distribution in our subsequent derivation. This is a standard result, but it is useful to include it here. We consider and exponential distribution with parameter $\lambda>0$, such that $\e{\lambda}(t)=\lambda e^{-\lambda t}$ ($t\ge0$). The Laplace transform is obtained as follows 
\begin{align}
\cL\left[\e{\lambda}(t)\right]
&= \int_{0^-}^\infty \lambda e^{-\lambda t} e^{-st}\,\dt \nonumber\\
&= \lambda\int_{0^-}^\infty e^{-(s+\lambda)t}\,\dt.
\end{align}
This integral is only convergent in the region $\Re(s)>-\lambda$. Within this region we have
\begin{equation}
\cL\left[\e{\lambda}(t)\right]=\frac{\lambda}{s+\lambda}.
\label{eq:app:laplace_exponential}
\end{equation}

\subsection{Laplace transform of an object $\delta(t)+z^{-1}\delta'(t)$}
We now show that the Laplace transform of $\delta(t)+z^{-1}\delta'(t)$ is $1+z^{-1}s$. The object $\delta'(t)$ is the derivative of the Dirac-$\delta$ distribution $\delta(t)$ (see Appendix~\ref{app:Background}\,\ref{app:Dirac} above) and $z>0$ is a constant. We have
\begin{align}
\cL\left[\delta(t)+z^{-1}\delta'(t)\right]
&= \int_{0^-}^\infty [\delta(t)+z^{-1}\delta'(t)]e^{-st} \,\dt \nonumber\\
&= \int_{0^-}^\infty e^{-st}\delta(t) \,\dt +z^{-1}\left[e^{-st}\delta(t)\right]_{0^-}^\infty \nonumber\\
&\quad+z^{-1}s\int_{0^-}^\infty e^{-st}\delta(t) \,\dt \nonumber\\
&= 1+z^{-1}s.
\label{eq:app:laplace_delta}
\end{align}
This expression has no singularities, and thus the region of convergence in terms of $s$ is the entire complex plane. Note that we explicitly define the lower integration limit as $0^-$ to include the $\delta$ function in the integral, and we have used $\lim_{t\to0^-}\delta(t)=0$.

\subsection{Convolution of an exponential distribution with an object $\delta(t)+z^{-1}\delta'(t)$} \label{app:Conv}
The convolution of an exponential distribution $\e{\lambda}(t)$ with an object of the form $\delta(t)+z^{-1}\delta'(t)$ (for constant $z$), as shown in Eq.~\eqref{eq:expconv}, is
\begin{align}
\e{\lambda} * &\left(\delta + z^{-1}\delta'\right) \nonumber\\
&= \int_0^\infty \lambda e^{-\lambda \tau}  \left[\delta(t-\tau)+z^{-1}\delta'(t-\tau)\right] \,\dtau \nonumber\\
&= \lambda e^{-\lambda t} +\lambda z^{-1}\delta(t) -\lambda^2 z^{-1} e^{-\lambda t} \nonumber\\
&= \frac{\lambda}{z}\delta(t) +\left(1-\frac{\lambda}{z}\right)\e{\lambda}(t).
\label{eq:app:convolution1}
\end{align}

\bigskip

\section{Calculation of absorption time distributions via Laplace transforms} \label{app:Calculate}
\subsection{Laplace representation}
As mentioned in the main text it is convenient to focus on the states $i=1,\dots,N-1$ of the birth-death process shown in Fig. 1 of the main paper. The dynamics of these states is given by $\vec{p}=\mathbb{A}\cdot \vec{p}$, where $\mathbb{A}$ is an $(N-1)\times (N-1)$ matrix, and where $p_i(t)$ is the probability that the system is in state $i$ at time $t$. We note that this is not a probability-conserving master equation, as probability mass continuously leaks into the absorbing states. We will use the notation $p_i$ (lower case) when we discuss the restricted system (with $\sum_{i=1}^{N-1}p_i(t)\leq 1$), and we will write $P_i(t)$ (upper case) when we discuss the full system, $i=0,\dots,N$. For the latter one always has $\sum_{i=0}^N P_i(t)=1$ at all times.

The formal solution of the equation $\vec{p}=\mathbb{A}\cdot \vec{p}$, restricted to sites $i=1,\dots,N-1$, reads
\begin{equation}
\vec{p}(t)=\exp(\mathbb{A}t)\cdot\vec{p}(0).
\end{equation}

We can take the Laplace transform and write the matrix exponential in the resolvent form
\begin{equation}
\widehat{\vec{p}}(s) = (s\id-\mathbb{A})^{-1} \cdot \vec{p}(0).
\end{equation}
We have here written $\widehat{\vec{p}}(s)$ for the Laplace transform $\cL[\vec{p}(t)]$. We consider initial conditions of the form $p_i(0)=\delta_{i,i_0}$ ($1\le i_0 \le N-1$). Our strategy is to compute $p_1(t)$ and $p_{N-1}(t)$, and from these the rates $d_1p_1(t)$ and $b_{N-1}p_{N-1}(t)$, with which probability arrives in the absorbing states, can be obtained. Thus we are interested in 
\begin{align}
\widehat{p}_{1|i_0}(s) &= \bigl[(s\id-\mathbb{A})^{-1}\bigr]_{1,i_0}, \nonumber\\
\widehat{p}_{N-1|i_0}(s) &= \bigl[(s\id-\mathbb{A})^{-1}\bigr]_{N-1,i_0}.
\end{align}

The $(i,j)$th element of the inverse of an invertible matrix $\mathbb{B}$ is given by $[\mathbb{B}^{-1}]_{ij}=C_{j,i}/\det \mathbb{B}$, where $C_{j,i}$ is the $(j,i)$th co-factor of $\mathbb{B}$. Thus we can write
\begin{equation}
\widehat{p}_{1|i_0}(s)  = \bigl[(s\id-\mathbb{A})^{-1}\bigr]_{1,i_0} = \frac{1}{\det(s\id-\mathbb{A})}C_{i_0,1},
\end{equation}
and likewise for the $(N-1,i_0)$th element. 

\subsection{Calculation of co-factors}
The co-factor $C_{i_0,1}$ is found from dropping column 1 and row $i_0$ from $s\id-\mathbb{A}$ and evaluating
\begin{widetext}
\begin{equation}
C_{i_0,1}=(-1)^{i_0+1}
\begin{vmatrix}
-d_2  & 0 \\
s+a_2 & -d_3 & 0 \\
-b_2 & s+a_3 & -d_4 & 0 \\
& \ddots & \ddots & \ddots & \ddots \\
& 0 &-b_{i_0-2} & s+a_{i_0-1} & -d_{i_0} &0\\
& & 0 & 0 & -b_{i_0} & s+a_{i_0+1} & -d_{i_0+2} &0\\
& & & \ddots & \ddots & \ddots & \ddots &\ddots \\
& & & & 0 & 0 & -b_{N-2} & s+a_{N-1}
\end{vmatrix},
\end{equation}
where $a_i=b_i+d_i$ is introduced for compactness. Using Laplace's formula, this can be written as
\begin{align}
C_{i_0,1} &= (-1)^{i_0+1}\left(\prod_{i=2}^{i_0}-d_i\right)
\begin{vmatrix}
s+a_{i_0+1} & -d_{i_0+2} & 0\\
-b_{i_0+1} & s+a_{i_0+2} & -d_{i_0+3} & 0\\
& \ddots & \ddots & \ddots & \\
& 0 & -b_{N-3} & s+a_{N-2} & -d_{N-1}\\
& & 0 & -b_{N-2} & s+a_{N-1}
\end{vmatrix}\nonumber\\
&= \left(\prod_{i=2}^{i_0} d_i\right)~ \det\left(s\id-\mathbb{A}_{(N-i_0-1)}\right) \nonumber\\
&= \prod_{i=2}^{i_0} d_i~ \prod_{\alpha=1}^{N-i_0-1}(s+x_\alpha).
\end{align}
\end{widetext}
The matrix $\mathbb{A}_{(N-i_0-1)}$ consists of the rows and columns $i_0+1,\dots,N-1$ of the matrix $\mathbb{A}$, i.e., it is the bottom right $(N-i_0-1)\times(N-i_0-1)$ sub-matrix of $\mathbb{A}$. The matrix $-\mathbb{A}_{(N-i_0-1)}$ has eigenvalues $x_\alpha>0$ ($\alpha=1,\dots,N-i_0-1$) and determinant $\det \left(-\mathbb{A}_{(N-i_0-1)}\right)=\chi_{i_0}=\prod_{\alpha=1}^{N-i_0-1} x_\alpha$.

\medskip

For the $(i_0,N-1)$th cofactor we have
\begin{widetext}
\begin{align}
C_{i_0,N-1} &= (-1)^{i_0+N-1}\left(\prod_{i=i_0}^{N-2}-b_i\right)
\begin{vmatrix}
s+a_1 & -d_2 & 0\\
-b_1 & s+a_2 & -d_3 & 0\\
& \ddots & \ddots & \ddots & \\
& 0 & -b_{i_0-3} & s+a_{i_0-2} & -d_{i_0-1}\\
& & 0 & -b_{i_0-2} & s+a_{i_0-1}
\end{vmatrix}\nonumber\\
&= \left(\prod_{i=i_0}^{N-2} b_i\right)~ \det\left(s\id-\mathbb{A}^{(i_0-1)}\right) \nonumber\\
&= \prod_{i=i_0}^{N-2} b_i~ \prod_{\alpha=1}^{i_0-1}(s+y_\alpha).
\end{align}
\end{widetext}
The matrix $\mathbb{A}^{(i_0-1)}$ consists of rows and columns $1,\dots,i_0-1$ of the matrix $\mathbb{A}$, i.e., it is the top-left $(i_0-1)\times(i_0-1)$ sub-matrix of $\mathbb{A}$. The matrix $-\mathbb{A}^{(i_0-1)}$ has eigenvalues $y_\alpha>0$ ($\alpha=1,\dots,i_0-1$) and determinant $\det \left(-\mathbb{A}^{(i_0-1)}\right)=\psi_{i_0}=\prod_{\alpha=1}^{i_0-1} y_\alpha$.

\subsection{Arrival time distributions}
Putting things together we have
\begin{align}
\widehat{p}_{1|i_0}(s) &=  \prod_{i=2}^{i_0} d_i \prod_{\alpha=1}^{N-i_0-1}(s+x_\alpha) \prod_{\beta=1}^{N-1}\frac{1}{s+\lambda_\beta},\nonumber\\
\widehat{p}_{N-1|i_0}(s) &=  \prod_{i=i_0}^{N-2} b_i \prod_{\alpha=1}^{i_0-1}(s+y_\alpha) \prod_{\beta=1}^{N-1}\frac{1}{s+\lambda_\beta}.
\end{align}
Here we have used $\det(s\id-\mathbb{A})=\prod_{\beta=1}^{N-1}{(s+\lambda_\beta)}$, where $\lambda_\beta>0$ are the eigenvalues of $-\mathbb{A}$. Using $\widehat{\dot{P}}_0(s)=d_1\widehat{p}_1(s)$ and $\widehat{\dot{P}}_N(s)=b_{N-1}\widehat{p}_{N-1}(s)$, we obtain the Laplace transforms of the absorption time distributions at sites $i=0$ and $i=N$, respectively. They are given by
\begin{align}
\widehat{\dot{P}}_{0|i_0}(s) &=  \prod_{i=1}^{i_0} d_i \prod_{\alpha=1}^{N-i_0-1}(s+x_\alpha) \prod_{\beta=1}^{N-1}\frac{1}{s+\lambda_\beta},\nonumber\\
\widehat{\dot{P}}_{N|i_0}(s) &=  \prod_{i=i_0}^{N-1} b_i\prod_{\alpha=1}^{i_0-1}(s+y_\alpha) \prod_{\beta=1}^{N-1}\frac{1}{s+\lambda_\beta}.
\label{eq:app:laplace_solution}
\end{align}

\subsection{Time representation of solutions} \label{app:Sol}
Combining Eqs.~\eqref{eq:app:laplace_exponential}, \eqref{eq:app:laplace_delta}, and \eqref{eq:app:laplace_solution}, we can write
\begin{widetext}
\begin{align}
\widehat{\dot{P}}_{0|i_0}(s) &=  \prod_{i=1}^{i_0} d_i \prod_{\alpha=1}^{N-i_0-1}\cL\bigl[x_\alpha\delta(t)+\delta'(t)\bigr] \prod_{\beta=1}^{N-1}\frac{\cL\bigl[\e{\lambda_\beta}(t)\bigr]}{\lambda_\beta},\nonumber\\
\widehat{\dot{P}}_{N|i_0}(s) &=  \prod_{i=i_0}^{N-1} b_i\prod_{\alpha=1}^{i_0-1}\cL\bigl[y_\alpha\delta(t)+\delta'(t)\bigr]\prod_{\beta=1}^{N-1}\frac{\cL\bigl[\e{\lambda_\beta}(t)\bigr]}{\lambda_\beta}.
\label{eq:app:laplace_solution2}
\end{align}
Using the fact that $\cL^{-1}\bigl[\widehat{f}(s)\cdot\widehat{g}(s)\bigr]=f * g$, we can perform the inverse Laplace transform of the expressions in Eq.~\eqref{eq:app:laplace_solution2}. We find
\begin{align}
\dot{P}_{0|i_0}(t) &=  \left[\frac{\prod\limits_{i=1}^{i_0} d_i  \prod\limits_{\alpha=1}^{N-i_0-1} x_\alpha}{\prod\limits_{\beta=1}^{N-1} \lambda_\beta}\right]
\e{\lambda_1} * \dots * \e{\lambda_{N-1}} * 
\left(\delta+x_1^{-1}\delta'\right) * \dots * \left(\delta+x_{N-i_0-1}^{-1}\delta'\right),\nonumber\\
\dot{P}_{N|i_0}(t) &=  \left[\frac{\prod\limits_{i=i_0}^{N-1} b_i  \prod\limits_{\alpha=1}^{i_0-1} y_\alpha}{\prod\limits_{\beta=1}^{N-1}\lambda_\beta}\right]
\e{\lambda_1} * \dots * \e{\lambda_{N-1}} *
\left(\delta+y_1^{-1}\delta'\right) * \dots * \left(\delta+y_{i_0-1}^{-1}\delta'\right).
\label{eq:app:general_solutions}
\end{align}
\end{widetext}
This is the expression shown in Eq.~\eqref{eq:gen_result}. We identify the prefactors in square brackets as the fixation probabilities $\phi_{0|i_0}$ and $\phi_{N|i_0}$, respectively.

\subsection{Symmetry of the fixation time distributions}
By choosing the initial conditions $i_0=N-1$ and $i_0=1$, the expressions in Eq.~\eqref{eq:app:general_solutions} can be reduced to
\begin{align}
\dot{P}_{0|N-1}(t)
&=  \left[\prod_{i=1}^{N-1}\frac{d_i}{\lambda_i}\right] \e{\lambda_1} * \dots * \e{\lambda_{N-1}} \nonumber\\
&= \phi_{0|N-1} E_{N-1}, \nonumber\\
\dot{P}_{N|1}(t)
&=  \left[\prod_{i=1}^{N-1}\frac{b_i}{\lambda_i}\right] \e{\lambda_1} * \dots * \e{\lambda_{N-1}} \nonumber\\
&= \phi_{N|1} E_{N-1},
\label{eq:app:antal}
\end{align}
where $E_\ell=\e{\lambda_1} * \dots * \e{\lambda_\ell}$. From this we conclude
\begin{equation}
\frac{\dot{P}_{0|N-1}(t)}{\phi_{0|N-1}} = \frac{\dot{P}_{N|1}(t)}{\phi_{N|1}},
\end{equation}
that is the conditional arrival time distribution at state $i=0$ given $i_0=N-1$ is equal to the conditional arrival time distribution at state $i=N$ given $i_0=1$. This symmetry has been known for the mean fixation time \cite{antal2006fixation,taylor2006symmetry}, and it was recently shown that the correspondence holds for the full distribution \cite{leier2014exact}.
Our approach offers an alternative way to obtain this intriguing result.

\bigskip

\section{Computing eigenspace representations} \label{app:Compute}
As the convolution operator is commutative, we can order the convolutions in Eq.~\eqref{eq:app:general_solutions} in any way. The exponential distributions and the objects of the form $\delta(t)+z^{-1}\delta'(t)$ in Eq.~\eqref{eq:app:general_solutions} can hence be combined in multiple ways.

\subsection{Convolutions I: Pairing $\delta+z_\alpha^{-1}\delta'$ with individual exponential distributions} \label{app:Pairing}
In this section we choose to couple $\e{\lambda_{N-\alpha}}(t)$ with $\delta(t)+x_\alpha^{-1}\delta'(t)$ for the purposes of $\dot{P}_{0|i_0}(t)$, and with $\delta(t)+y_\alpha^{-1}\delta'(t)$ when we calculate $\dot{P}_{N|i_0}(t)$. Carrying out these convolutions in Eq.~\eqref{eq:app:general_solutions}, we arrive at 
\begin{widetext}
\begin{align}
\dot{P}_{0|i_0}(t) &= \phi_{0|i_0}\times
\e{\lambda_1} * \dots * \e{\lambda_{i_0}} * \nonumber \\
&\qquad \left[\frac{\lambda_{i_0+1}}{x_{N-i_0-1}}\delta+\left(1-\frac{\lambda_{i_0+1}}{x_{N-i_0-1}}\right)\e{\lambda_{i_0+1}}\right] * \dots * \left[\frac{\lambda_{N-1}}{x_1}\delta+\left(1-\frac{\lambda_{N-1}}{x_1}\right)\e{\lambda_{N-1}}\right], \nonumber\\
\dot{P}_{N|i_0}(t) &= \phi_{N|i_0}\times
\e{\lambda_1} * \dots * \e{\lambda_{N-i_0}} * \nonumber \\
&\qquad \left[\frac{\lambda_{N-i_0+1}}{y_{i_0-1}}\delta+\left(1-\frac{\lambda_{N-i_0+1}}{y_{i_0-1}}\right)\e{\lambda_{N-i_0+1}}\right] * \dots *
\left[\frac{\lambda_{N-1}}{y_1}\delta+\left(1-\frac{\lambda_{N-1}}{y_1}\right)\e{\lambda_{N-1}}\right].
\end{align}
\end{widetext}

We stress that the objects $\delta(t)+x_\alpha^{-1}\delta'(t)$ [or $\delta(t)+y_\alpha^{-1}\delta'(t)$] can be paired with any of the exponential distributions. We chose to match these at the end of the exponential chain so that the reduced chains can be systematically compared. This is the representation described in Fig.~\ref{fig:fig2}(a).

\subsection{Convolutions II: Recursively convolving with exponential chain} \label{app:Matching}
If we write Eqs.~\eqref{eq:app:general_solutions} in the form
\begin{align}
\frac{\dot P_{0|i_0}(t)}{\phi_{0|i_0}} &= E_{N-1} * \left(\delta+x_1^{-1}\delta'\right) * \dots * \left(\delta+x_{N-i_0-1}^{-1}\delta'\right), \nonumber \\
\frac{\dot P_{N|i_0}(t)}{\phi_{N|i_0}} &= E_{N-1} * \left(\delta+y_1^{-1}\delta'\right) * \dots * \left(\delta+y_{i_0-1}^{-1}\delta'\right),
\end{align}
where $E_\ell=\e{\lambda_1} * \dots * \e{\lambda_\ell}$, then we can recursively convolute the objects involving $\delta$ functions onto the chain of exponentials from the right. We note that
\begin{equation}
E_\kappa * \left(\delta+z^{-1}\delta'\right) = \left[\frac{\lambda_\kappa}{z} E_{\kappa-1}(t)+\left(1-\frac{\lambda_\kappa}{z}\right)E_\kappa(t)\right],
\end{equation}
which follows directly from Eq.~\eqref{eq:app:convolution1}.
From this, each of the recursive convolutions introduces a new exponential chain with one step less. For example,
\begin{widetext}
\begin{align}
\frac{\dot P_{0|i_0}(t)}{\phi_{0|i_0}} &= E_{N-1} * \left(\delta+x_1^{-1}\delta'\right) * \dots * \left(\delta+x_{N-i_0-1}^{-1}\delta'\right) \nonumber\\
&=
\left[\left(1-\frac{\lambda_{N-1}}{x_1}\right)E_{N-1} +\frac{\lambda_{N-1}}{x_1}E_{N-2}\right] 
* \left(\delta+x_2^{-1}\delta'\right) * \dots * \left(\delta+x_{N-i_0-1}^{-1}\delta'\right)\nonumber\\
&=
\Biggl\{\left(1-\frac{\lambda_{N-1}}{x_1}\right)\left(1-\frac{\lambda_{N-1}}{x_2}\right)E_{N-1}+\left[\left(1-\frac{\lambda_{N-1}}{x_1}\right)\frac{\lambda_{N-1}}{x_2} + \frac{\lambda_{N-1}}{x_1}\left(1-\frac{\lambda_{N-2}}{x_2}\right)\right]E_{N-2}\nonumber\\
&\hspace{.5cm}+\frac{\lambda_{N-1}}{x_1}\frac{\lambda_{N-2}}{x_2} E_{N-3}\Biggr\}
* \left(\delta+x_3^{-1}\delta'\right) * \dots * \left(\delta+x_{N-i_0-1}^{-1}\delta'\right).\nonumber\\
\end{align}

Performing all the convolutions leads to the following expressions:
\begin{align}
\frac{\dot{P}_{0|i_0}(t)}{\phi_{0|i_0}} &= \left(\prod_{\alpha=1}^{N-i_0-1}\frac{1}{x_\alpha}\right)
\sum_{\alpha=1}^{N-i_0}\Biggl[E_{N-\alpha}(t) \left(\prod_{\beta=1}^{\alpha-1} \lambda_{N-\beta}\right) \times\nonumber\\
&\hspace{0cm}\sum_{j_1=1}^{\alpha}\left(x_{j_1}-\lambda_{N-j_1}\right)
\sum_{j_2=j_1}^{\alpha}\left(x_{j_2+1}-\lambda_{N-j_2}\right)
\sum_{j_3=j_2}^{\alpha}~~ \dots
\sum_{\substack{j_{N-i_0-\alpha}=\\j_{N-i_0-\alpha-1}}}^{\alpha}\left(x_{j_{N-i_0-\alpha}+N-i_0-\alpha-1}-\lambda_{N-j_{N-i_0-\alpha}}\right)
\Biggr],\nonumber\\
\frac{\dot{P}_{N|i_0}(t)}{\phi_{N|i_0}} &= \left(\prod_{\alpha=1}^{i_0-1} \frac{1}{y_\alpha}\right)
\sum_{\alpha=1}^{i_0}\Biggl[E_{N-\alpha}(t) \left(\prod_{\beta=1}^{\alpha-1} \lambda_{N-\beta}\right) \times\nonumber\\
&\hspace{0cm}\sum_{j_1=1}^{\alpha}\left(y_{j_1}-\lambda_{N-j_1}\right)
\sum_{j_2=j_1}^{\alpha}\left(y_{j_2+1}-\lambda_{N-j_2}\right)
\sum_{j_3=j_2}^{\alpha}~~ \dots
\sum_{\substack{j_{i_0-\alpha}=\\j_{i_0-\alpha-1}}}^{\alpha}\left(y_{j_{i_0-\alpha}+i_0-\alpha-1}-\lambda_{N-j_{i_0-\alpha}}\right)
\Biggr].
\label{eq:app:convolution2}
\end{align}
\end{widetext}

These expressions can be written as
\begin{align}
\dot{P}_{0|i_0}(t) &= \phi_{0|i_0}\sum_{\alpha=1}^{N-i_0} G^{(x)}_{N-\alpha} E_{N-\alpha}(t), \nonumber\\
\dot{P}_{N|i_0}(t) &= \phi_{N|i_0}\sum_{\alpha=1}^{i_0} G^{(y)}_{N-\alpha} E_{N-\alpha}(t),
\label{eq:app:sums}
\end{align}
where $G_{N-\alpha}^{(x)}$ and $G_{N-\alpha}^{(y)}$ are constants (independent of time).
The fixation time distributions are thus linear combinations of the distributions $E_{N-\alpha}$. We note here the equalities
\begin{equation}
\sum_{\alpha=1}^{N-i_0} G^{(x)}_{N-\alpha} =1,~~\sum_{\alpha=1}^{i_0} G^{(y)}_{N-\alpha}=1.
\end{equation}

We now proceed to express the above linear combination of exponential chains [Eq.~\eqref{eq:app:sums}] as the single chain shown in Fig.~\ref{fig:fig2}(b). In the schematic shown in this figure the system can transition to two possible states if currently in eigenstate $\alpha$: either $\alpha\to\alpha+1$ or $\alpha\to N$. These paths have transition rate $F_\alpha\lambda_\alpha$ and $(1-F_\alpha)\lambda_\alpha$, respectively. The total rate of transitioning out of $\alpha$ is then $\lambda_\alpha$, and the waiting time at $\alpha$ is an exponential distribution with parameter $\lambda_\alpha$, no matter whether the system transitions to $\alpha+1$ or to $N$. The quantity $F_\alpha$ denotes the probability that the next state of dynamics in eigenspace is $\alpha+1$, if the system is currently in eigenstate $\alpha$. With probability $1-F_\alpha$ the next state is eigenstate $N$.  Evaluating the probability of a trajectory in terms of $F_\alpha$, and then matching with Eq.~\eqref{eq:app:sums} gives 
\begin{equation}
F_\alpha = \frac{1-\sum\limits_{\kappa=1}^\alpha G_\kappa}{1-\sum\limits_{\kappa=1}^{\alpha-1}G_\kappa} ~~\mbox{for}~\alpha<N-1.
\end{equation}

We can express all transition rates in Fig.~\ref{fig:fig2}(b) as
\begin{equation}
T_{\alpha\to\beta} = \left\{
\begin{matrix}
F_\alpha \lambda_\alpha, & \beta=\alpha+1<N, \\
(1-F_\alpha)\lambda_\alpha, & \alpha\le N-1,~\beta=N.
\end{matrix}
\right.
\end{equation}

\subsection{Evaluation of `bottom-line' arrival time distributions} \label{app:BottomLine}
The final expressions for the (un-normalized) arrival time distributions follow directly from Eq.~\eqref{eq:app:sums}. First we note that the convolution of $\ell$ exponential distributions has the form
\begin{equation}
E_\ell(t) = \left(\prod_{\alpha=1}^\ell \lambda_\alpha\right) \sum_{\alpha=1}^{\ell} \prod_{\substack{\beta=1\\\beta\ne \alpha}}^\ell \frac{1}{\lambda_\beta-\lambda_\alpha} e^{-\lambda_\alpha t}.
\end{equation}

Substituting this expression into Eq.~\eqref{eq:app:sums}, or equivalently Eq.~\eqref{eq:app:convolution2}, we arrive at the final expressions
\begin{align}
\dot{P}_{0|i_0}(t) &= \left(\prod_{\ell=1}^{i_0}d_\ell\right) \sum_{\alpha=1}^{N-1}\left[ \frac{\prod\limits_{\gamma=1}^{N-i_0-1} \left(x_\gamma-\lambda_\alpha\right)}{\prod\limits_{\substack{\beta=1\\\beta\ne \alpha}}^{N-1}(\lambda_\beta-\lambda_\alpha)}e^{-\lambda_\alpha t}\right], \nonumber\\
\dot{P}_{N|i_0}(t) &= \left(\prod_{\ell=i_0}^{N-1}b_\ell\right) \sum_{\alpha=1}^{N-1}\left[ \frac{\prod\limits_{\gamma=1}^{i_0-1} \left(y_\gamma-\lambda_\alpha\right)}{\prod\limits_{\substack{\beta=1\\\beta\ne \alpha}}^{N-1}(\lambda_\beta-\lambda_\alpha)}e^{-\lambda_\alpha t}\right].
\label{eq:app:bottom_line}
\end{align}
Evaluating these expressions requires the calculation of the eigenvalues $x_{\alpha}, y_{\alpha}$, and $\lambda_{\alpha}$.

\subsection{Accuracy and efficiency of simulation-based approaches} \label{app:Accuracy}
In the inset of Fig.~\ref{fig:fig3} we show that computing arrival time distributions directly from simulations is less efficient than computing them exactly using our formalism. To generate the CPU-time curve for the simulation method we measure the distance, $d$, between the histogram of arrival times at state $N$, $\rho_N$, against the exact distribution. This distance is defined by
\begin{equation}
d = \frac{1}{2}\sum_{i=0}^M \left|(t_{i+1}-t_i)\rho_N(t_i) -  \int_{t_i}^{t_{i+1}} \frac{\dot{P}_{N|i_0}}{\phi_{N|i_0}} \,\dt \right|,
\end{equation}
where $M$ is the number of histogram bins. Note that this is the continuous analog of the distance measure between distributions to be discussed in the next section. The histogram, $\rho_N$, is populated with an increasing amount of independent realizations until the distance falls below $d=1/2$. We then plot the time taken to complete this number of simulation runs of the given process.

\bigskip
\section{Relation to equilibration processes} \label{app:Equilibration}

\subsection{Dynamics without mutation} \label{app:NoMutation}
In the system without mutation all realizations reach fixation eventually.  If the dynamics is started from state $i_0$ [i.e., $P_i(t=0)=\delta_{i,i_0}$] the stationary state of the birth-death process, i.e., the terminal distribution, is of the form
\begin{equation}
\Phi_{i|i_0}=(1-\phi_{N|i_0}) \delta_{i,0}+\phi_{N|i_0}\delta_{i,N}, ~~(i=0,\dots,N).
\end{equation}
The quantity $\phi_{N|i_0}$ is the probability that the process reaches the absorbing state $N$. The probability of being absorbed at $0$ is $1-\phi_{N|i_0}$.

Let us now consider the distance of the distribution $\vec{P}(t)$ from this distribution $\vec{\Phi}=(1-\phi_{N|i_0},0,\dots,0,\phi_{N|i_0})$.  In line with the existing literature \cite{levin2009markov,black2012mixing} we use the distance measure 
\begin{equation}
d[\vec{P},\vec{Q}]=\frac{1}{2}\sum_{i=0}^N \left|P_i-Q_i\right|,
\end{equation}
for two distributions $\vec{P}$ and $\vec{Q}$. We then have
\begin{align}
d[\vec{P}(t),\vec{\Phi}] &= \frac{1}{2}\Biggl[\left|P_0(t)-(1-\phi_{N|i_0})\right| \nonumber\\
&\quad+ \sum_{i=1}^{N-1} P_i(t) + \left|P_N(t)-\phi_{N|i_0}\right|\Biggr].
\end{align}
Probability continuously flows into the absorbing states, hence $P_0(t)\leq 1-\phi_{N|i_0}$ and $P_N(t)\leq \phi_{N|i_0}$ [$P_0(t)$ approaches $1-\phi_{N|i_0}$ from below with time, and similarly for $P_N(t)$]. We can therefore simplify the above expression, and we are left with
\begin{align}
d[\vec{P}(t),\vec{\Phi}]
&= \frac{1}{2}\left[1-P_0(t)-P_N(t)+\sum_{i=1}^{N-1}P_i(t)\right]\nonumber\\
&= 1-P_0(t)-P_N(t).
\label{eq:app:fixdistance}
\end{align}
This means that the distance $d(t)=d[\vec{P}(t),\vec{\Phi}]$ is given by the probability that the system has not yet reached fixation in either of the absorbing states by time $t$. This in turn means that $1-d(t)=\mbox{Pr}(t_{\rm fix}\leq t)$ is the probability to have reached fixation by time $t$, i.e., it is the cumulative distribution of the unconditional fixation time $t_{\rm fix}$. The quantity $-\dot{d}(t)$ is therefore the probability density function of the unconditional fixation time.

As a side remark we note that the mean unconditional fixation time can be expressed as follows
\begin{align}
\avg{t_{\rm fix}}
&= \int_0^\infty \left[-\dot d(t)\right] t \,\dt \nonumber \\
&= \bigl[-d(t)t\bigr]_0^\infty+ \int_0^\infty  d(t) \,\dt \nonumber \\
&= \int_0^\infty d(t) \,\dt.
\end{align}
Thus the mean unconditional fixation time is the area under the curve $d(t)$.

\subsection{Dynamics with mutation} \label{app:Mutation}
\subsubsection{Definitions}
We now consider systems with mutation, which occurs with rate $u\ll1$. This means that the states $0$ and $N$ are no longer absorbing. More specifically we consider systems in which $b_0=\order{u}$ and $d_N=\order{u}$, i.e., escape from the states $0$ and $N$ occurs with a rate proportional to $u$. All remaining transition rates ($b_i, d_i$, $i=1,\dots,N-1$) are $\order{u^0}$, and hence are only affected at sub-leading order by the introduction of mutation. For $u=0$, one recovers the case with absorbing states ($b_0=d_N=0$). Below we will compare the system with small mutation rates with the system without mutation.

The rates used for the analysis shown in Figs.~\ref{fig:fig4}, \ref{fig:fig5}, and \ref{fig:fig6} are given by (see Ref.~\cite{black2012mixing})
\begin{align}
b_i &= (1-u)\frac{i(N-i)}{N} g[+\Delta\pi(i)] + \frac{u}{2}\frac{(N-i)^2}{N},\nonumber \\
d_i &= (1-u)\frac{i(N-i)}{N} g[-\Delta\pi(i)] + \frac{u}{2}\frac{i^2}{N}.
\label{eq:app:mutation}
\end{align}

\subsubsection{Stationary state}
For $u>0$ there are no absorbing states and the dynamics reaches a stationary distribution, $\vec{P}^{\rm st}$, with full support, i.e., $P_i^{\rm st}>0$ for all $i=0,1,\dots, N$. This distribution can be expressed as \cite{vankampen}
\begin{align}
P_{i>0}^{\rm st} &= \left(\prod_{j=1}^{i}\frac{b_{j-1}}{d_j}\right)P_0^{\rm st}, \nonumber\\
P_0^{\rm st} &= \left(1+\sum_{i=1}^N \prod_{j=1}^i \frac{b_{j-1}}{d_j}\right)^{-1}.
\label{eq:app:iter}
\end{align}
In the limit of small mutation rates ($0< u N \ll 1$), it can be seen that
\begin{equation}
\frac{P_i^{\rm st}}{P_0^{\rm st}} = \frac{b_0}{b_i} \left(\prod_{j=1}^{i}\frac{b_j}{d_j}\right) = \order{u} ~\mbox{for}~ i=1,\dots,N-1.
\end{equation}
Together with the normalization condition ($\sum_{i=0}^N P_i^{\rm st}=1$) we can determine that $P_0^{\rm st}$ and $P_N^{\rm st}$ must be $\order{u^0}$, and the remaining probability masses are $\order{u}$. With this we can write
\begin{equation}
P_i^{\rm st}=(1-\sigma)\delta_{i,0}+\sigma\delta_{i,N}+\order{u},
\end{equation}
for $i=0,\dots,N$, where $\sigma=\order{u^0}$. This indicates that in the limit of small mutation the distribution is peaked at the boundaries.

\subsubsection{Intermediate distribution}
The system without mutation ($u=0$) has terminal distribution $\vec{\Phi}$, as discussed above. In particular $P_0(t)$ and $P_N(t)$, the probability masses concentrated in the absorbing states, grow with time and we have $P_0(t) \to 1-\phi_{N|i_0}$ and $P_N(t) \to \phi_{N|i_0}$ as $t\to \infty$.

The rates of the system with small mutation differ from those of the system without mutation only by corrections of $\order{u}$. At small mutation rates we expect the dynamics on a short timescale ($t\ll u^{-1}$) to be essentially that of the system without mutation, the effects of mutation only set in at a longer time. Of course the boundary states are no longer absorbing, but we argue that the system initially approaches a distribution close to $\vec{\Phi}$ before reaching its stationary distribution $\vec{P}^{\rm st}$.

This can be seen mathematically as follows:

\noindent Let $\vec{P}^{(u=0)}(t)=[P^{(u=0)}_0(t),\dots, P^{(u=0)}_N(t)]$ be the probability distribution of the system without mutation. The time evolution is described by the master equation
\begin{equation}
\dot{\vec{P}}^{(u=0)} = \mathbb{M} \cdot \vec{P}^{(u=0)},
\end{equation}
where $\mathbb{M}$ is an $(N+1)\times(N+1)$ transition matrix (to be distinguished from the truncated matrix $\mathbb{A}$). Let $\vec{P}^{(u)}(t)$ be the distribution in the system with mutation whose evolution is described by
\begin{equation}
\dot{\vec{P}}^{(u)} = \bigl(\mathbb{M}+u\mathbb{Q}\bigr) \cdot \vec{P}^{(u)},
\end{equation}
where the matrix $\mathbb{Q}$ reflects the difference between the systems with and without mutation. The elements of both matrices $\mathbb{M}$ and $\mathbb{Q}$ are $\order{u^0}$. Now, let $\vec{q}(t)=\vec{P}^{(u)}(t)-\vec{P}^{(u=0)}(t)$, such that
\begin{equation}
\dot{\vec{q}} = \mathbb{M} \cdot \vec{q}+u\mathbb{Q} \cdot \vec{P}^{(u)}.
\label{eq:app:qdot}
\end{equation}
We want to calculate how the separation, $\vec{q}$, grows in time given that both systems (with and without mutation) start from the same initial condition [i.e., $\vec{q}(0)=\vec{0}$]. For this purpose it is convenient to work in the eigenspace of $\mathbb{M}$. This matrix has two zero eigenvalues, $\mu_0=\mu_1=0$, with eigenvectors $(\vec{v}_0)_i=\delta_{i,0}$ and $(\vec{v}_1)_i=\delta_{i,N}$. These are the absorbing states of the system without mutation.

Decomposing $\vec{q}(t)=\sum_\alpha \widetilde{q}(t)_\alpha \vec{v}_\alpha$ into eigendirections $\vec{v}_\alpha$ of $\mathbb{M}$ we have
\begin{equation}
\dot{\widetilde{q}}_\alpha = \mu_\alpha \widetilde{q}_\alpha + u g_\alpha(t),
\label{eq:app:qalphadot}
\end{equation}
where we have written $\mathbb{Q}\cdot\vec{P}^{(u)}(t)=\sum_\alpha g_\alpha(t)\vec{v}_\alpha$ and we note that $g_\alpha(t)=\order{u^0}$. This can be integrated to give
\begin{equation}
\widetilde{q}_\alpha(t)=ue^{\mu_\alpha t}\int_0^t e^{-\mu_\alpha \tau}g_\alpha(\tau)~\dtau.
\label{SM:eq:exp}
\end{equation}

On short time scales ($t\ll u^{-1}$) $\widetilde{\vec{q}}(t)=\order{u}$, and hence the separation $\vec{q}(t)$ is also $\order{u}$. That is to say in the limit $u\to0$, both systems (with and without mutation) initially evolve along the same trajectory. On this time scale both systems approach the distribution $\vec{\Phi}$.

On a longer time scale [$t=\order{u^{-1}}$], differences between the two systems become of $\order{u^0}$. However these differences are concentrated on the states $i=0$ and $i=N$. Effectively, a redistribution of probability mass between the boundary states takes place. The distribution of the system with mutation evolves from $\Phi_{i|i_0}=(1-\phi_{N|i_0})\delta_{i,0}+\phi_{N|i_0}\delta_{i,N}$ to $P^{\rm st}_i=(1-\sigma)\delta_{i,0}+\sigma\delta_{i,N}$, as shown in Fig.~\ref{fig:fig5}.

\subsubsection{Distance from stationarity and mixing times}
Approximating the stationary distribution of the system with small mutation rate as $P^{\rm st}_i=(1-\sigma)\delta_{i,0}+\sigma\delta_{i,N}$ we have
\begin{align}
d[\vec{P}^{(u)}(t),\vec{P}^{\rm st}] &\approx \frac{1}{2}\Biggl[\left|P_0^{(u)}(t)-(1-\sigma)\right| \nonumber\\
&\quad+ \sum_{i=1}^{N-1} P_i^{(u)}(t) + \left|P_N^{(u)}(t)-\sigma\right|\Biggr],
\label{eq:app:distu}
\end{align}
for the distance between the distribution $\vec{P}^{(u)}(t)$ of the system with mutation at time $t$ and the stationary distribution. While $P_0(t)$ and $P_N(t)$ are monotonically increasing with time in the system without mutation, this is not necessarily the case if there is mutation. Hence we cannot easily drop the absolute values in Eq.~\eqref{eq:app:distu} as in the case without mutation.

We observe, though, that $P_0^{(u)}(t=0)=0$ and $P_N^{(u)}(t=0)=0$ for $0<i_0<N$. Hence there is an initial phase of the dynamics in which $P_0^{(u)}(t)<1-\sigma$ and $P_N^{(u)}(t)<\sigma$. Let us write $t^*$ for the first time at which either $P_0(t^*)=1-\sigma$ or $P_N(t^*)=\sigma$ (whichever happens first). Prior to this time we have
\begin{align}
d[\vec{P}^{(u)}(t),\vec{P}^{\rm st}]
&\approx \frac{1}{2}\Biggl[(1-\sigma)-P_0^{(u)}(t) \nonumber\\
&\quad+ \sum_{i=1}^{N-1}P_i^{(u)}(t) + \sigma-P_N^{(u)}(t)\Biggr]\nonumber \\
&= 1-P_0^{(u)}(t)-P_N^{(u)}(t).
\end{align}
This is the same as the distance to the fixation distribution, $\vec{\Phi}$, in the system without mutation, given in Eq.~\eqref{eq:app:fixdistance}.
From this we can conclude that
\begin{equation}
d[\vec{P}^{(u)}(t),\vec{P}^{\rm st}] = d[\vec{P}^{(u)}(t),\vec{\Phi}] ~~\mbox{for}~t<t^*.
\end{equation}
This is illustrated in Figs.~\ref{fig:fig4} and \ref{fig:fig6}, where the distributions $\vec{P}^{\rm st}$ and $\vec{\Phi}$ are represented as single points in the $(P_0,P_N)$ plane.

Therefore, for times $t<t^*$, we have the relation
\begin{align}
d[\vec{P}^{(u)}(t),\vec{P}^{\rm st}] &= d[\vec{P}^{(u)}(t),\vec{\Phi}] \nonumber\\
&\approx d[\vec{P}^{(u=0)}(t),\vec{\Phi}] = \Pr(t_{\rm fix}>t).
\label{eq:app:equivalence}
\end{align}
That is to say, the mixing time is related to the cumulative distribution of fixation times. The first time at which $d[\vec{P}^{(u)}(t),\vec{P}^{\rm st}]=\varepsilon$, provided $t<t^*$, is approximately the $(1-\varepsilon)$th percentile of the unconditional fixation time distribution. Choosing $\varepsilon=1/2$, we have the equivalence with between the mixing time and the median fixation time. This equivalence holds for the example shown in Figs.~\ref{fig:fig4} and \ref{fig:fig5}. However, it does not hold in the example shown in Fig.~\ref{fig:fig6} where the stationary state is asymmetric in the $(P_0,P_N)$ plane. In this case the equivalence in Eq.~\eqref{eq:app:equivalence} breaks down prior to the time at which $d[\vec{P}^{(u)}(t),\vec{P}^{\rm st}]=1/2$.

\begin{acknowledgments}
P.A. gratefully acknowledges support from the EPSRC (United Kingdom).
\end{acknowledgments}

%

\end{document}